\documentclass[pre,twocolumn,groupedaddress,showkeys,showpacs]{revtex4}

\usepackage{graphicx}
\usepackage{CJK}
\begin{document}


\title{Densest columnar structures of hard spheres from sequential deposition}


\author{Ho-Kei Chan}


\affiliation{Foams and Complex Systems, School of Physics, Trinity College Dublin, College Green, Dublin 2, Ireland }
\email{epkeiyeah@yahoo.com.hk}


\date{\today}
\begin{abstract}
\indent
The rich variety of densest columnar structures of identical hard spheres inside a cylinder can surprisingly be constructed from a simple and computationally fast sequential deposition of cylinder-touching spheres, if the cylinder-to-sphere diameter ratio is $D\in[1,2.7013]$. This provides a direction for theoretically deriving \emph{all} these densest structures and for constructing such densest packings with nano-, micro-, colloidal or charged particles, which all self-assemble like hard spheres.
\end{abstract}

\pacs{62.23.St, 81.16.Dn, 81.16.Rf}



\maketitle

\section {Introduction}
\indent
The problem of finding the densest configurations of identical hard spheres inside a cylinder is related to many different aspects in science and engineering: Similar densest structures (e.g. zigzags and helices) have in recent years been observed for the self-assembly of nano-spheres inside carbon nanotubes \cite{Khlobystov_2004}, charged particles on cylindrical surfaces \cite{Srebnik_2011}, micro-spheres inside micro-channels \cite{Vanapalli_2008} and colloidal particles inside cylindrical channels \cite{Lohr_2010}, and also for the formation of colloidal crystal wires \cite{Tymczenko_2008}, the reason being that in all such cases the particles close-pack themselves like hard spheres. Relevant problems of interest include (i) finding a theoretical derivation of the corresponding densest packings of hard spheres and (ii) finding a general methodology for constructing them in practice.

For this problem, a wide range of chiral and achiral densest structures for a diameter ratio of $D \equiv$ cylinder diameter / sphere diameter $\in[1,1 + 1/sin(\pi/5)]$ have previously been found by simulated annealing \cite{Pickett_2000,Mughal_2011,Hodak_2003} and by other computational means \cite{Duran_2009,Koga_2006}, where $1 + 1/sin(\pi/5)=2.7013$ is a critical value of $D$ below which the densest configurations are entirely made up of cylinder-touching spheres. In particular, the densest structures for $D\in[2,2.7013]$, which have their entire structural information contained on the cylinder surface, can be described by a phyllotactic scheme \cite{Mughal_2011} that is relevant to the study of plant morphology.

\begin{figure}[htbp]
\begin{center}\includegraphics[width=7.2cm,height=5.4cm]{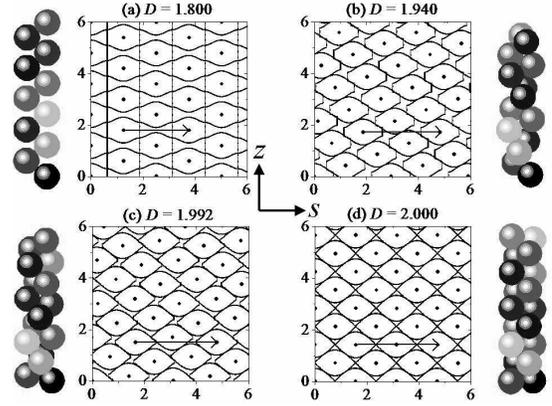}
\caption{The densest structures and their phyllotactic diagrams for various cases of $D\in(1,2]$: (a) the achiral planar zigzag at $D=1.800$, (b) the chiral single helix at $D=1.940$, (c) the chiral double helix at $D=1.992$, and (d) the achiral doublets, which have two spheres located on one $z$ position, at $D=2.000$. For each case, the required periodicity $|\textbf{V}|=(D-1)\pi$ along $s$ is indicated by a solid arrow.
}\label{phyllo_D_leq_2}
\end{center}
\end{figure}

In this article, it is shown that $all$ the densest structures for $D\in[1,2.7013]$ can surprisingly be constructed from a simple and computationally fast sequential deposition of cylinder-touching spheres which maximizes the number of spheres per unit length everywhere along the cylinder axis (typically less than 15 minutes of computational time for each value of $D$; much faster than simulated annealing which sometimes can take up a whole week \cite{Mughal_2011_comm}). As demonstrated in Fig. \ref{phyllo_D_leq_2} for $D\in(1,2]$, the first four phases constructed from such a deposition procedure are in complete agreement with the planar zigzag, single helix, double helix and achiral doublets predicted from simulated annealing \cite{Pickett_2000,Mughal_2011}. The findings show that all these densest packings can be derived theoretically by means of maximizing the abovementioned number density of spheres, and establish sequential deposition as a possible methodology for constructing such densest packings with the aforementioned physical systems.

\begin{figure}[htbp]
\begin{center}\includegraphics[width=6.6667cm,height=5cm]{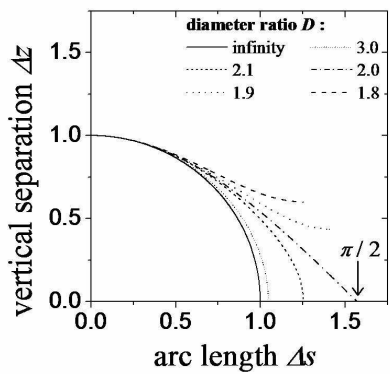}
\caption{Plot of the vertical separation $\Delta z$ against the arc length $\Delta s$ for spheres in contact at various values of the diameter ratio $D$. The regimes $D\in[1,2)$ and $D \geq 2$ correspond to the absence and presence of a solution for $\Delta z = 0$, respectively. As $D\rightarrow \infty$, the relation between $\Delta z$ and $\Delta s$ approaches $(\Delta z)^2 + (\Delta s)^2=1$ for the 2-D packing of circular discs.
}\label{contact_eq}
\end{center}
\end{figure}

This 3-D problem of packing identical spheres onto the inner surface of a cylinder is equivalent to a problem of packing objects in 2-D space, and is related to the 2-D problem of packing circular discs: In cylindrical polar coordinates $(\rho,\phi,z)$ with a radial position $\rho = (D-1)/2$ and an arc length $s = (D-1)\phi/2$, where the length quantities $\rho$, $z$ and $s$ are all expressed in units of the sphere diameter, the equation for two such spheres in contact is given by
\begin{equation}\label{eq_1}
(\Delta z)^{2}+\frac{{(D-1)}^{2}}{2}(1-cos{\Delta \phi}) = 1,
\end{equation}
which involves only two position variables (i.e. the differences in $z$ and $\phi$ between the two spheres in contact), corresponding to a 2-D problem, and can be derived easily using trigonometric identities. This implies the following (Fig. \ref{contact_eq}): (i) For $D\in[1,2)$, the cylinder's surface is so curved that no two spheres can be accommodated on the same $z$ position, i.e no solution exists for $\Delta z = 0$. (ii) As $D\rightarrow\infty$, the cylindrical surface becomes flat and the equation for spheres in contact approaches that for the 2-D packing of circular discs; In the densest configuration, the sphere centres form a perfect triangular lattice on the phyllotactic plot \cite{Mughal_2011} of $z$ against $s$ such that every sphere is in contact with 6 other spheres. (iii) For any finite value of $D$, the problem of finding an analytic solution to the densest configuration is complicated by the requirement of a periodicity $|\textbf{V}| = (D-1)\pi$ along the $s$ axis \cite{Mughal_2011} and by the relatively complex mathematical form of Eq. (\ref{eq_1}). Finding an analytic solution for cases of finite $D$ thus remains a challenging theoretical problem.

To gain insights into such an analytic solution, a semi-analytic approach is adopted here, based on the following conjecture: For any given value of $D$, if the packing of spheres is locally densest everywhere in the columnar structure, the overall structure would be at its globally densest configuration. For a sufficiently long column of spheres, let $\Delta N/\Delta z$ be the average number density of spheres (number of spheres per unit length) along $z$, where $N$ are the indices of the spheres in ascending order of their $z$ positions. This number density is related to the volume fraction, $V_{F} \equiv$ volume of spheres / volume of cylinder, as follows:
\begin{equation}\label{eq_2}
V_{F}=\frac{2}{3D^2}\frac{\Delta N}{\Delta z}.
\end{equation}
Thus the densest columnar structure, which by definition has the greatest volume fraction, also corresponds to the densest packing of spheres along $z$. It follows, from the abovementioned conjecture, that the densest columnar structure can be constructed from a sequential deposition of individual spheres onto the lowest possible $z$ positions (determined across the entire $2\pi$ range of $\phi$) at which they would come into contact with one or more other spheres. This is because such an iterative process would ensure the packing of spheres along $z$ to be densest possible at every $z$ location. A template comprised of spheres, which covers the entire $2\pi$ range of $\phi$, is needed as a base for the initial deposition of spheres. For $D\in[1,2]$, Eq. (\ref{eq_1}) for spheres in contact is valid for any value of $\Delta\phi$ so that, among other possible choices of a multi-sphere template, a single sphere is already enough to make up such a template. For $D > 2$, however, this equation is valid only for a limited range of $\Delta\phi$, which satisfies the condition $cos\Delta\phi\in[1-2/(D-1)^2,1]$, so that a template comprised of more than one sphere is necessary. In general, for any given value of $D$, the columnar structures of spheres generated from the abovementioned deposition procedure depend on their underlying templates, and the densest structure can be obtained by varying across a diversity of templates.

\begin{figure}[htbp]
\begin{center}\includegraphics[width=7.2cm,height=5.4cm]{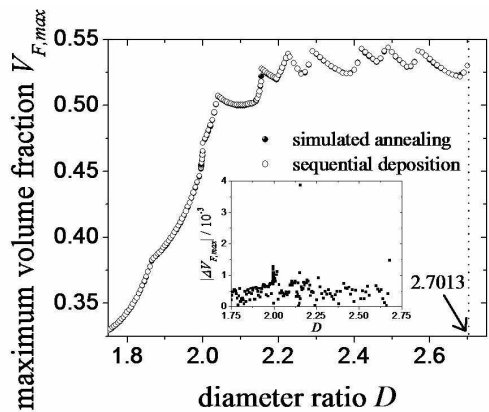}
\caption{Plot of the maximum volume fraction $V_{F,max}$ against the diameter ratio $D$, showing a complete overlap of data points between this deposition algorithm and simulated annealing (data supplied by Mughal \cite{Mughal_2011_comm}). The inset shows that the differences in $V_{F,max}$ between the two algorithms are of the order of $10^{-3}$, i.e. less than 1\% of percentage differences, and thus comparable to the numerical uncertainties in the simulations. The occasional peaks in $|\Delta V_{F,max}|$ are due to the relatively larger numerical uncertainties in the proximity of a structural transition.
}\label{max_VF_vs_D}
\end{center}
\end{figure}

The different templates are generated through a deposition scheme in which the template structures depend only on the self-assigned relative positions $(\Delta\phi)_{2,1}=\phi_{2}-\phi_{1}\in[0,\pi]$ and $(\Delta z)_{2,1}=z_{2}-z_{1}\in[0,1]$ between the two initially placed spheres: After the first sphere is placed at an arbitrary position on the plane of $(s,z)$ coordinates, a particular value of $(\Delta\phi)_{2,1}$ is chosen for the second sphere, where the value of $(\Delta z)_{2,1}$ is given by the corresponding solution to Eq. (\ref{eq_1}). If such a solution does not exist, the value of $(\Delta z)_{2,1}$ is assigned separately. Along the same deposition direction of increasing or decreasing $\phi$, which determines the chirality of a chiral structure, additional spheres are deposited if the existing template as made up of these two initially filled spheres does not cover the $2\pi$ range of $\phi$. Inspired by the relevant triangular-lattice-like phyllotactic patterns \cite{Mughal_2011}, the positions of such additional spheres are chosen in the spirit to maintain a continuity in the relative positions $\Delta\phi_{N,N-1}=\phi_{N}-\phi_{N-1}$ and $\Delta z_{N,N-1}=z_{N}-z_{N-1}$ of successively filled spheres: For $N>2$, the angular position of the $N^{th}$ sphere is set to be $(\Delta\phi)_{N,1}=\phi_{N}-\phi_{1}=(N-1)(\Delta\phi)_{2,1}$, and the sphere is deposited in a $z$ position at which it comes into contact with one or more other spheres. If such a $z$ position does not exist, the sphere is placed at $(\Delta z)_{N,1}=z_{N}-z_{1}=(N-1)(\Delta z)_{2,1}$. Deposition of such additional spheres stops whenever the template created has covered the $2\pi$ range of $\phi$. The iterative process of sequential deposition of spheres then starts from the angular position of the last deposited sphere in the process of template creation: In each iterative step, a sphere is deposited onto the lowest possible $z$ position at which it would come into contact with one or more other spheres. Starting from the angular position of the previously filled sphere, this lowest $z$ position is determined via a $2\pi$ scan along the same deposition direction of increasing or decreasing $\phi$ as that for the process of template creation. Should there be degeneracy where this lowest $z$ position occurs at more than one angular position, the first such angular position encountered in the $2\pi$ scan is chosen. After a sufficiently long (e.g. 10 times of the sphere diameter) column of spheres has been generated, the average number density $\Delta N/\Delta z$ is evaluated via a linear fit of $N$ against $z$, and the volume fraction is computed using Eq. (\ref{eq_2}). The abovementioned procedure is repeated for a continuous range of $(\Delta\phi)_{2,1}$ from 0 to $\pi$, and for a continuous range of $(\Delta z)_{2,1}$ from 0 to 1 whenever the value of $(\Delta z)_{2,1}$ has to be assigned separately. As such, a quantitative agreement in the maximum volume fraction $V_{F,max}$ between simulated annealing and this deposition algorithm has been obtained for the whole range of $D\in[1,2.7013]$, which is demonstrated for $D\in[1.750,2.7013]$ in Fig. \ref{max_VF_vs_D} (The numerical resolution of $D$ is set to be 0.001, except for a better resolution near $D=2.7013$).

\begin{figure}[htbp]
\begin{center}\includegraphics[width=7.2cm,height=5.4cm]{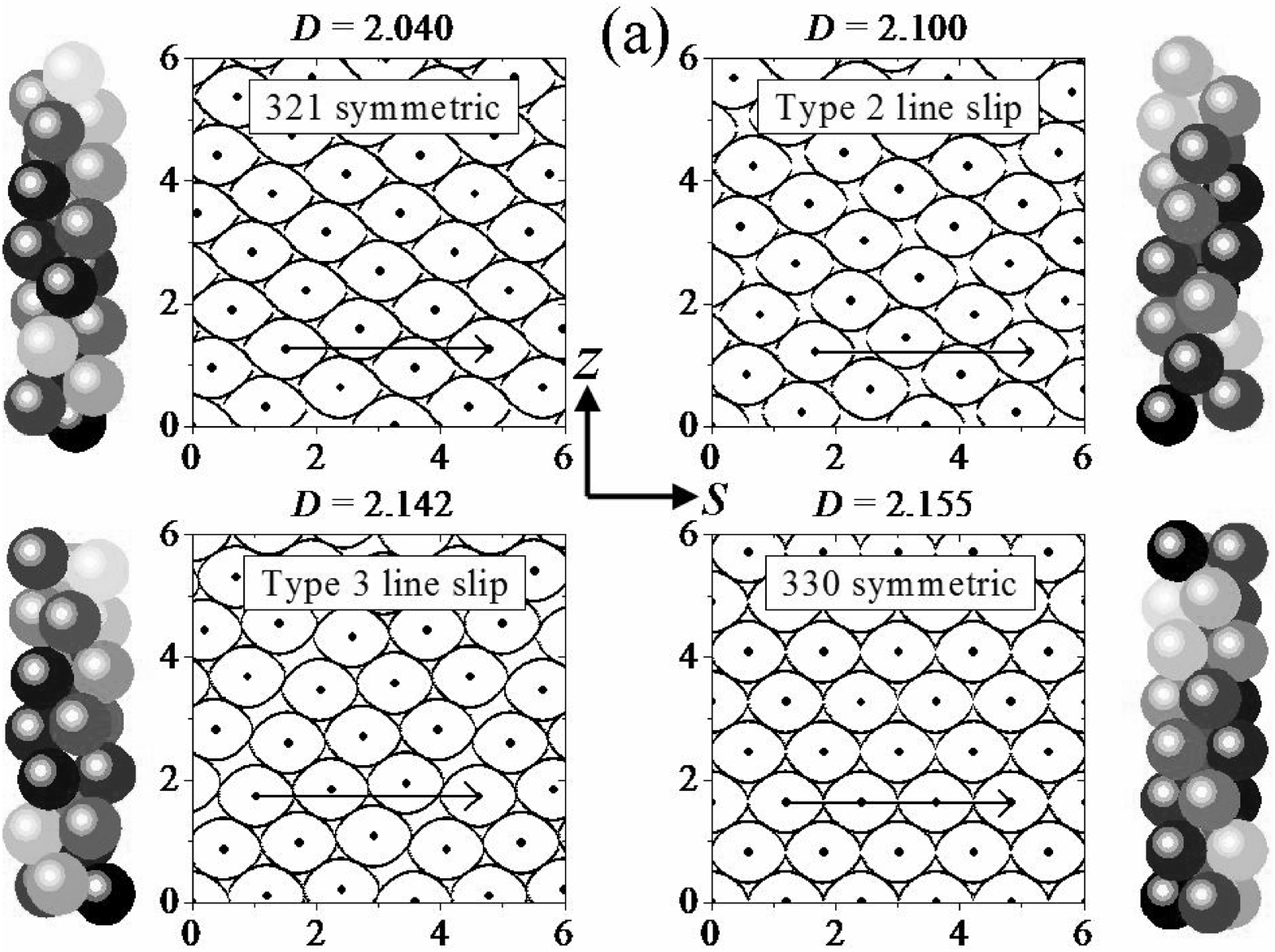}
\includegraphics[width=7.2cm,height=2.7cm]{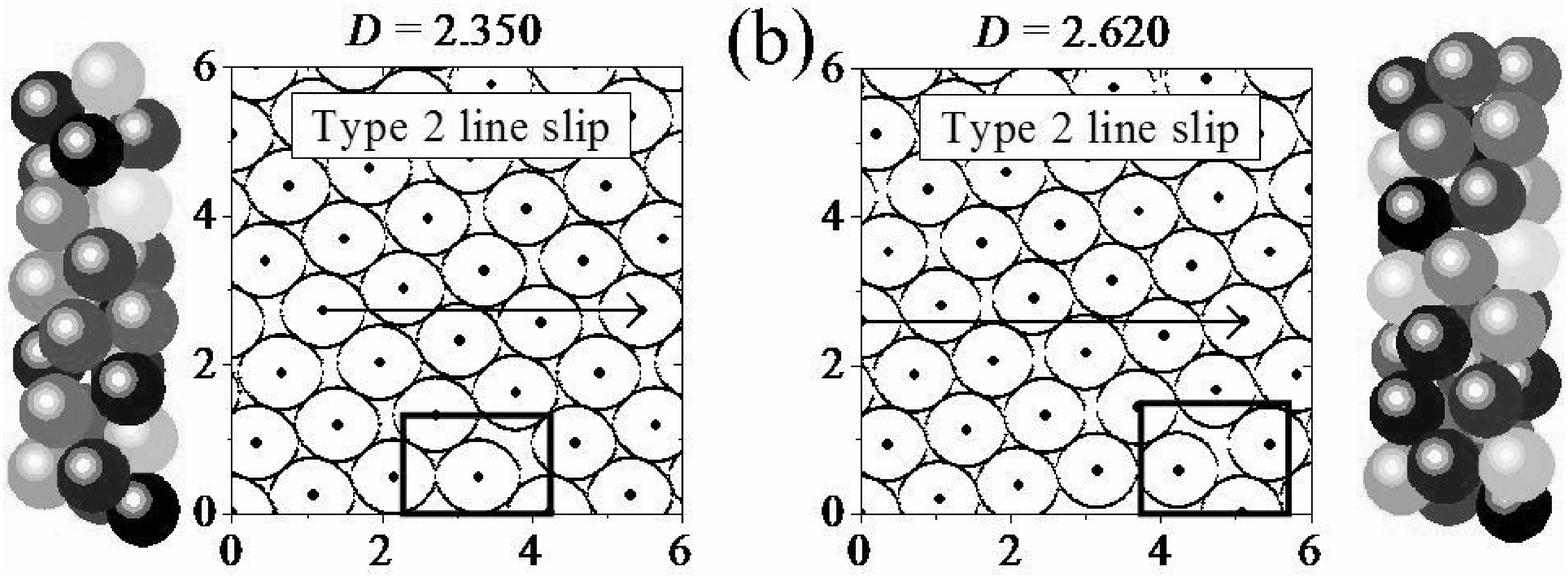}
\caption{(a) Sequence of structures as $D$ increases from 2.040 to 2.155: 321 symmetric at $D=2.040$ $\rightarrow$ Type 2 line slip, e.g. at $D=2.010$ $\rightarrow$ Type 3 line slip, e.g. at $D=2.142$ $\rightarrow$ 330 symmetric at $D=2.155$. (b) Two examples of periodicity breaking at $D=2.350$ and $D=2.620$; in each case, the region that does not belong to the periodic structure is highlighted with a rectangle.
}\label{phyllo_D_geq_2}
\end{center}
\end{figure}

\begin{figure}[htbp]
\begin{center}\includegraphics[width=6.667cm,height=5.0cm]{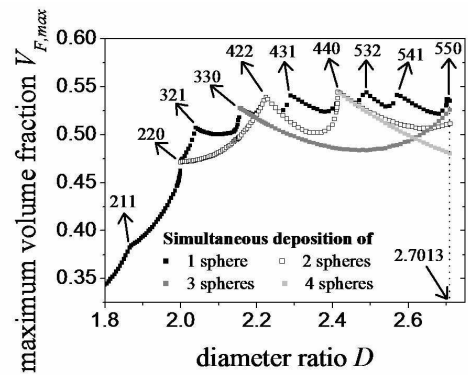}
\caption{Simultaneous deposition of $u=1,2,3$ and $4$ spheres at regular spacings of $\phi$ in each deposition step. All the metastable curves for lower-density packings correspond to cases of $u>1$. The symmetric structures are indicated along with their phyllotactic notations.
}\label{simul_depos}
\end{center}
\end{figure}

A complete agreement in the predicted densest structures has also been obtained for $D\in[1,2.7013]$ between the two algorithms. For $D\in(1.000,2.000]$ the following sequence of densest structures has previously been predicted from simulated annealing \cite{Pickett_2000,Mughal_2011}: (i) the achiral planar zigzag for $D\in(1.000,1.866]$, (ii) the chiral single helix for $D\in(1.866,1.990]$, (iii) the chiral double helix for $D\in(1.990,2.000)$, and (iv) the achiral doublets, which have two spheres located on one $z$ position, at $D=2.000$. Here, the deposition algorithm yields the same sequence of densest structures for the same regimes of $D$. A gallery of the corresponding phyllotactic diagrams is shown in Fig. \ref{phyllo_D_leq_2}. In creating such diagrams, the boundaries of the equivalent close-packed objects in 2-D space are plotted from an equation which comes from the following transformation of Eq. (\ref{contact_eq}): $\Delta z\rightarrow2\Delta z$ and $\Delta s\rightarrow2\Delta s$. This can be easily understood from the limiting case of $D\rightarrow\infty$: Applying such a transformation to $(\Delta z)^2+(\Delta s)^2=1$ for the 2-D packing of circular discs correctly yields the equation for a circle of radius = $1/2$. For $D\in[2.000,2.7013]$, the phyllotactic patterns from simulated annealing generally consist of regions with triads of interconnected spheres, which can be described in terms of a distorted triangular lattice \cite{Mughal_2011}, as well as line defects that arise due to the required periodicity $|\textbf{V}|=(D-1)\pi$ along $s$. These are referred to as \emph{line-slip} structures \cite{Mughal_2011}. At certain values of $|\textbf{V}|$ or equivalently $D$, line defects are absent so that on a phyllotactic diagram every sphere appears to be in contact with 6 other spheres. Such defect-free structures, which generally correspond to the density maxima in Fig. (\ref{max_VF_vs_D}), are described as \emph{symmetric} \cite{Mughal_2011}. In such cases, the periodicity vector is given by $\textbf{V}=c_{1}\textbf{b}_{1}+c_{2}\textbf{b}_{2}$ where $\textbf{b}_{1}$ and $\textbf{b}_{2}$ are any two of the three basis vectors. The coefficients $c_{1}\geq0$ and $c_{2}\geq0$ are by definition chosen from the indices in the phyllotactic notation $lmn$ \cite{Mughal_2011}, where $l\equiv m+n$. An example of such symmetric structures is the 220 structure in Fig. \ref{phyllo_D_leq_2}(d). For this regime of $D$, the densest structures predicted from the deposition algorithm are in complete agreement with the line-slip and symmetric structures predicted from simulated annealing. For example, as shown in Fig. \ref{phyllo_D_geq_2}(a), the deposition algorithm yields the same sequence of structures as $D$ increases from 2.040 to 2.155 \cite{Pickett_2000,Mughal_2011}: 321 symmetric at $D=2.040$ $\rightarrow$ type 2 line slip for $D\in(2.040,2.142)$ $\rightarrow$ type 3 line slip for $D\in(2.142,2.155)$ $\rightarrow$ 330 symmetric at $D=2.155$, where the different types of line-slip structures are defined in Ref. \cite{Mughal_2011}. In some cases a periodic line-slip structure comes into existence only after a few spheres have been deposited; Two examples of such periodicity breaking are demonstrated in Fig. \ref{phyllo_D_geq_2}(b). Lower-density packings, which might also form a subject of interest, can be generated from a simultaneous deposition of multiple spheres at each deposition step (Fig. \ref{simul_depos}). For $D>2.7013$, the structures with cylinder-touching spheres become hollow so that the aforementioned agreement with the 3-D densest structures from simulated annealing no longer exists. At $D\rightarrow \infty$, the cylinder's surface becomes flat and essentially a 2-D triangular lattice of spheres is obtained. A sequential stacking of such 2-D layers along their perpendicular direction in AB or ABC sequence yields respectively the densest 3-D packings of HCP or FCC ($V_{F}=0.74048$, the difference of which with the peak values in Figs. \ref{max_VF_vs_D} and \ref{simul_depos} clearly reflects the boundary effects imposed by the cylinder). For low values of $D$ the experimental observations of the planar zigzag \cite{Khlobystov_2004,Vanapalli_2008,Lohr_2010,Tymczenko_2008,Hodak_2003}, the helical phases \cite{Khlobystov_2004,Lohr_2010,Tymczenko_2008,Hodak_2003} and the achiral doublets \cite{Khlobystov_2004,Tymczenko_2008,Hodak_2003} could be related to the following: (i) The particles self-assemble like hard spheres, in the sense that their shapes remain approximately spherical. (ii) During the process of structural formation, the particles are driven by a uni-directional force (e.g. gravity) which helps to maximize the number density $\Delta N/\Delta z$. In the future a more general deposition scheme that also applies to the regime of $D>2.7013$ and to the packing of binary hard spheres \cite{Hopkins_2011,Brouwers_2007} should be developed. Possibilities of applying the current algorithm for the synthesis of chiral molecules should also be explored.

This work was funded by IRCSET. Stimulating discussions with Denis Weaire, Stefan Hutzler and Adil Mughal, as well as indispensable support from Chloe Po-Yee Wong, are gratefully acknowledged.

\end{document}